\documentclass[conference]{IEEEtran}
\IEEEoverridecommandlockouts
\usepackage{multirow}
\usepackage{cite}
\usepackage{amsmath,amssymb,amsfonts}
\usepackage{algorithmic}
\usepackage{graphicx}
\usepackage{textcomp}
\usepackage{xcolor}
\usepackage{mathtools}
\usepackage{commath}
\usepackage{hyperref}
\hypersetup{colorlinks,allcolors=black}
\usepackage[compact]{titlesec}
\titlespacing{\section}{0pt}{2ex}{1ex}
\titlespacing{\subsection}{0pt}{1ex}{0ex}
\titlespacing{\subsubsection}{0pt}{0.5ex}{0ex}
\usepackage{amsmath}

\def\BibTeX{{\rm B\kern-.05em{\sc i\kern-.025em b}\kern-.08em
    T\kern-.1667em\lower.7ex\hbox{E}\kern-.125emX}}
    
\begin{document}
 
\title{A Scalable Deep Learning Framework for Multi-rate CSI Feedback under Variable Antenna Ports}

\author{Yu-Chien Lin, Ta-Sung Lee, and Zhi Ding
\thanks{Y.-C Lin is with the Department of Electrical and Computer Engineering,
University of California at Davis, Davis, CA, USA, and
was affiliated with National Yang Ming Chiao Tung University, Taiwan (e-mail: ycmlin@ucdavis.edu).

Z. Ding is with the Department of Electrical and Computer Engineering,
University of California, Davis, CA, USA (e-mail: zding@ucdavis.edu)

T.-S Lee is with the Institute of Communications Engineering, National Yang Ming Chiao Tung University, Taiwan (e-mail: tslee@mail.nctu.edu.tw).}

\thanks{This work is based on materials supported by the National Science Foundation under Grants 2029027 and 2002937 (Lin, Ding) and by the Center for Open Intelligent Connectivity under the Featured Areas Research Center Program within the framework of the Higher Education Sprout Project by the Ministry of Education (MOE) of Taiwan, and partially supported by the Ministry of Science and Technology (MOST) of Taiwan under grant MOST 111-2634-F-A49-009 and MOST MOST 110-2224-E-A49-001 (Lee and Lin).}}

\maketitle

\begin{abstract}

Channel state information (CSI) 
at transmitter
is crucial for massive MIMO downlink systems to achieve high spectrum and energy efficiency. Existing works
have provided deep learning
architectures for CSI feedback and recovery at the eNB/gNB 
by reducing user feedback overhead and improving  recovery accuracy. However,
existing DL architectures tend to be inflexible
and non-scalable as models are often
trained according to a preset number
of antennas
for a given compression ratio. 
In this work, we develop a flexible
and scalable learning frame work based
on a divide-and-conquer approach (DCA).
This new DCA architecture can flexibly accommodate different numbers of 3GPP antenna ports and dynamic
levels of feedback compression. 
Importantly, it also significantly reduces computational complexity and
memory size by allowing UEs to  feedback segmented downlink CSI. 
We further propose a multi-rate successive convolution encoder 
with fewer than 1000 parameters. 
Test results demonstrate superior performance, good scalability,
and low complexity for both
indoor and outdoor channels.

\end{abstract}


\section{Introduction}
Massive Multiple-input multiple-output (MIMO) technologies play an important role in improving spectrum and energy efficiency of 5G and future generation wireless networks. 
The strength of massive MIMO hinges on
accurate downlink channel state information (CSI) at the basestation or gNodeB (gNB).
Without the benefit of uplink/downlink (UL/DL) channel reciprocity assumed in time-division duplxing (TDD) systems, 
frequency-division duplexing (FDD) systems typically rely on user 
equipment (UE) feedbacks
to gNBs for DL CSI recovery. 
The growing
large number of DL transmit antennas envisioned in millimeter wave bands or higher frequencies
\cite{TULVCAN} requires a vast amount of uplink
feedback information and resources such 
bandwidth and power. To conserve bandwidth
and UE battery, efficient compression
of CSI feedback is vital to broad 
FDD deployment of massive MIMO technologies.

From radio physics, cellular CSI exhibits
limited delay spread (sparsity).
Efficient UE feedback can take advantage of sparsity for CSI compression. 
To leverage CSI sparsity for improving feedback efficiency, a deep autoencoder framework \cite{CsiNet} deployed encoders at UE and 
decoder at serving station (gNB) for CSI compression and recovery, respectively. This and other related works have demonstrated
the efficacy of CSI recovery
with the aids of deep learning autoencoder \cite{CsiNet+, CRNet, DeepCMC, CLNet}. 
Physical insights with respect to slow temporal variations of propagation scenarios, similar propagation conditions of nearby UEs, and similarity of UL/DL radio paths reveal significant temporal, spatial, and spectral CSI correlations 
respectively.
Beyond basic autoencoders, more recent works take advantages of various 
correlated channel information such as past CSI \cite{CsiNet+, MarkovNet}, CSI of nearby UEs \cite{CoCsiNet}, and UL CSI \cite{CQNET, DualNet, DualNet-MP} to aid and improve the recovery of DL CSI at base stations. Additional works considered antenna array geometry to explicitly exploit UL/DL angular reciprocity  \cite{Yacong_reciprocity, Xing_reciprocity}.

To extract underlying correlated antenna information, current CSI frameworks compress and recover DL CSI over all antennas. This leads to high model complexity because of the large input size.
There have attempts to directly reduce encoder's model complexity \cite{CsiNet-SM, KD, MD} with limited success. 
We consider the physical insight that since {\em geometric size of massive MIMO 
antenna array spans multiple wavelengths,
only nearby antennas exhibit non-negligible
CSI correlation}. According, our
low complexity framework focuses on
adjacent antennas (or antenna ports)
with stronger spatial correlation to
compress the CSI of
large array via a divide-and-conquer approach (DCA).

This works aims to systematically
simplify deep learning architecture and
and computational complexity for DL CSI
feedback while limiting the
accuracy loss of CSI recovery. We develop a flexible and scalable 
multi-rate CSI feedback framework along with a lightweight convolutional encoder. Our contributions are summarized below. 
\begin{itemize}
    \item The DCA framework CSI is a new learning-based compression and recovery mechanism that systematically reduces the model size by exploiting both
    strong and weak CSI
    correlations among
adjacent and non-adjacent antennas, respectively.
    
\item Our framework is flexible and
applicable to multi-rate CSI compression 
ratios, including a universal encoder for all compression ratios.

\item The new framework is scalable to large antenna sizes. 

\item The framework incorporates
a swapping encoder mechanism which 
generates a few extra feedback bits
to further improve the CSI recovery performance.
\end{itemize}



\section{System Model}
We consider a single-cell MIMO FDD link in which
a gNB using a $N_a$ antenna ports communicates with single antenna UEs. Following the 3GPP specification, pilot symbols (or DMRS) for each antenna port are uniformly placed in frequency domain for downlink transmission. Assuming each subband contains $N_f$ subcarriers with spacing of ${\Delta}f$ and a downsampling rate $\text{DR}_f$, the frequency interval between consecutive pilots is $\text{DR}_f{\cdot}{\Delta}f$. We denote $\mathbf{h}_{i} \in \mathbb{C}^{M_f \times 1}$ as DL pilot CSI of the $i$-th AP where $M_f$ is the number of pilots. By collecting DL pilot CSI of each AP, a pilot sampled DL CSI matrix $\widetilde{\mathbf{H}}$ is related to the full CSI matrix $\bar{\mathbf{H}}$ is given by
\begin{equation}
    \widetilde{\mathbf{H}} = \bar{\mathbf{H}}\mathbf{Q}_{\text{DR}_f} =
    \begin{bmatrix}
    \mathbf{h}_1^H\\
    \mathbf{h}_2^H\\
    \vdots\\
    \mathbf{h}_{N_a}^H
    \end{bmatrix} \in \mathbb{C}^{N_a \times M_f},
    \label{DL pilot CSI matrix}
\end{equation}
where $\mathbf{Q}_{\text{DR}_f}$ is a downsampling matrix with downsampling rate $\text{DR}_f$ and superscript $(\cdot)^H$ denotes conjugate transpose. 

To reduce feedback overhead, we
exploit the delay sparsity of DL CSI
and transform full DL CSI into delay domain by applying discrete Fourier transform (DFT) and conduct a truncation to discard those near-zero elements in the high delay region as follows:

\begin{equation}
    \mathbf{H} = \bar{\mathbf{H}} \cdot \mathbf{F}\cdot 
    \underbrace{\left[\begin{array}{c} \mathbf{I}_{N_t\times N_t}\\
\mathbf{0}\end{array}\right]}_{\mathbf{T}},
    \label{CSI in delay domain}
\end{equation}
where $\mathbf{F} \in \mathbb{C}^{M_f \times M_f}$ denotes a DFT matrix and ${M_f \times N_{t}}$ matrix
$\mathbf{T}$ performs
delay domain truncation that drop the $M_f-N_t$ trailing columns of $\widetilde{\mathbf{H}} \cdot \mathbf{F}$.

Autoencoder structure has been widely adopted in several CSI feedback frameworks. An encoder UE compresses the DL CSI for uplink feedback and a decoder at gNB recovers the DL truncated
time domain CSI according to the feedback from UE. Most research exploits convolutional layers to compress and recover the DL pilot CSI via
\begin{equation}
    \mathbf{q} = f_\text{en}(\mathbf{H}), 
    \label{original encoder}
\end{equation}
\begin{equation}
    \widehat{\mathbf{H}} = f_\text{de}(\mathbf{q}). \label{original decoder}
\end{equation}
\section{Multi-rate CSI Feedback Framework with Flexible Number of Antenna Ports}

There have been notable progresses in terms of recovery performance among the recent autoencoder-based CSI feedback frameworks \cite{DualNet-MP, CANet, CoCsiNet, ENet}. 
Since UEs have limited resources \cite{ENet}, an important consideration is
the reduction of complexity and required storage at UE. 
Unfortunately, na\"ively following the example
of autoencoders in image compression leads
to the directly input of full DL CSI matrix $\mathbf{H}$ as an image to deep learning
architecture to extract pixel-wise features. Because of the inevitably large input size, 
fat learning models at both UE and gNB make it
highly challenging to effectively reduce the
model complexity and storage need.  

This raises an question: is it necessary to feed full DL CSI matrix into the model for encoding the spatial features between different APs? The answer may not be yes. That is because most antenna configuration (half-wavelength antenna spacings are usually adopted) just satisfies the spatial Nyquist rate condition\footnote{As a rule of thumb, the channels of two antennas spaced with one wavelength are nearly independent.} (i.e., the spatial sampling rate is at the borderline of avoiding spatial aliasing effect). In other words, there are barely margins or correlation between antennas. Thus, it is unnecessary to extract the correlation between antennas. 

\begin{figure}
    \centering
    \resizebox{3.4in}{!}{
    \includegraphics{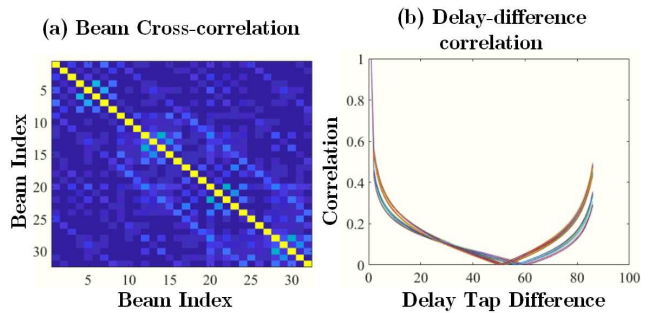}}
    \caption{(a) Cross-correlation between different beams (we consider a $8\times4$ orthogonal beam set), and (b) correlation versus various delay tap difference (we consider CSIs of $32$ antennas denoted by curves with different colors). The low cross-correlation between beams and the high similarity of these curves in delay domain imply the possibility to compress and recovery CSI antenna-by-antenna.}
    \label{fig: CorBtAntStaForAnt}
\end{figure}

We can get more insights from the following preliminary results. Figs. \ref{fig: CorBtAntStaForAnt} (a) and (b) show the correlation between different antennas and the statistics at different delay taps for different antennas. Obviously, there are weak correlation between antennas and the statistics at different delay taps for different antennas are similar. Thus, it is quite reasonable to use the same model to encode and decode the DL CSI for different APs.

\subsection{Divide-and-Conquer (DC) Framework}
Given the preliminary results of Figs. \ref{fig: CorBtAntStaForAnt} (a) and (b), the inter-antenna independence and similar statistics of delay profile of different antennas herald a simpler CSI feedback framework with a divide-and-conquer manner. In this section, we propose a DC framework which divde a full DL CSI into several pieces and then compress and recover them individually. 

We first define a new quantity, \textit{base number}, to be the dimension of the spatial domain of the new framework input. Assuming base number is $K$, as compared to the full DL CSI with size of $N_a \times N_f$, we first divide the full DL CSI matrix into $N_a/K$ secondary DL CSI matrices with the same size of $K \times N_f$ given as follows:
\begin{equation}
\mathbf{H}_{i} = \mathbf{P}_i\mathbf{H}, \ \forall {i = 1,2,...,N_a/K},   
\label{reduced-size DL CSI}
\end{equation}
\begin{equation}
\mathbf{P}_i = \begin{bmatrix}
\mathbf{e}_{K\cdot(i-1)+1}^T;\mathbf{e}_{K\cdot(i-1)+2}^T;...;\mathbf{e}_{K{\cdot}i}^T
\end{bmatrix}.    
\label{Sipping matrix}
\end{equation}

With the knowledge of similar statistics of real and imaginary parts of CSIs \cite{ENet}, we can further decompose the DL CSI into real and imaginary parts and feed them into an autoencoder network individually for DL pilot CSI compression and recovery which can be expressed as
\begin{equation}
    \mathbf{q}_{i}^{(R)}, \mathbf{q}_{i}^{(I)} = f_\text{en}(Real\{\mathbf{H}_{i}\}), f_\text{en}(Imag\{\mathbf{H}_{i}\}).
\end{equation}
\begin{equation}
    Real\{\widehat{\mathbf{H}}_{i}\}, Imag\{\widehat{\mathbf{H}}_{i}\}  = f_\text{de}(\mathbf{q}_{i}^{(R)}), f_\text{de}(\mathbf{q}_{i}^{(I)}).
\end{equation}
Note that the input size is reduced by a factor of $N_a/2K$. The full DL pilot CSI estimate is given by concatenating the estimates of the secondary DL CSI matrices via
\begin{equation}
    Real\{\widehat{\mathbf{H}}\} = \begin{bmatrix}
    Real\{\widehat{\mathbf{H}_{1}}\};Real\{\widehat{\mathbf{H}_{2}}\};...;Real\{\widehat{\mathbf{H}}_{{N_a/K}}\}
    \end{bmatrix}
\end{equation}
\begin{equation}
    Imag\{\widehat{\mathbf{H}}\} = \begin{bmatrix}
    Imag\{\widehat{\mathbf{H}_{1}}\};Imag\{\widehat{\mathbf{H}_{2}}\};...;Imag\{\widehat{\mathbf{H}}_{{N_a/K}}\}
    \end{bmatrix}
\end{equation}
\begin{equation}
    \widehat{\mathbf{H}} = Real\{\widehat{\mathbf{H}}\} + 1j\cdot Imag\{\widehat{\mathbf{H}}\}
\end{equation}

\subsection{Multi-rate Successive Convolutional CSI Feedback Framework}

\begin{figure}
    \centering
    \resizebox{3.4in}{!}{
    \includegraphics{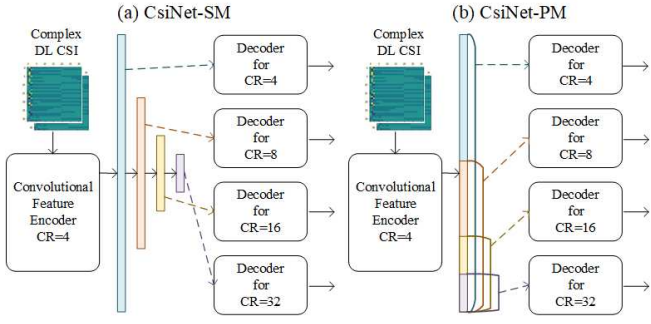}}
    \caption{Illustration of previous multi-rate CSI feedback frameworks, CsiNet-SM and CsiNet-PM. The encoders share model parameters at different compression ratios except for FC layers, which contribute the majority of model complexity.}
    \label{fig: CsiNet-SM}
\end{figure}

\begin{figure*}
    \centering
    \resizebox{6.8in}{!}{
    \includegraphics{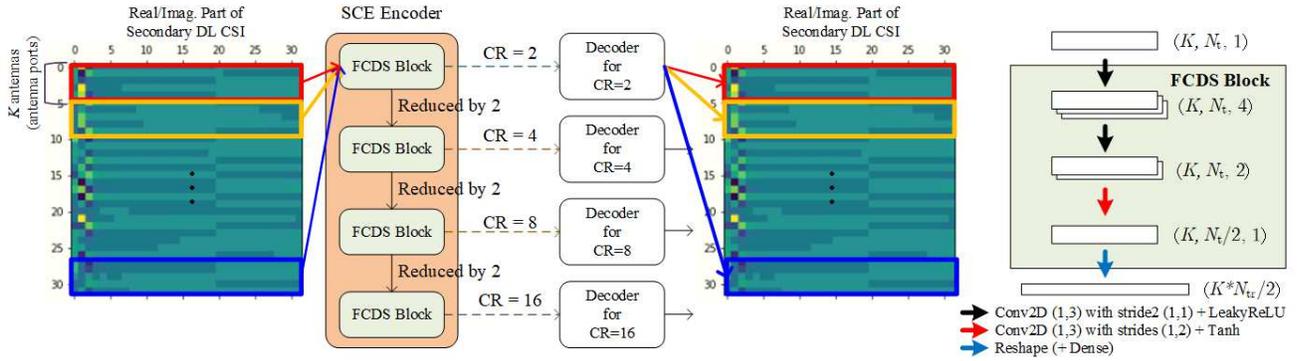}}
    \caption{DC Framework and SCE Network Architecture. The input is first divided into real and imaginary parts and separated into secondary CSI matrices. These matrices are parallelly fed into the SCE network and recovered after combining the estimates of secondary CSI matrices. Note that, at encoder, after each FCDS block, the total size of input is reduced by a factor of two and those fully convoluted FCDS blocks share parameters. We also provide another version of encoder, named as SCE-Dense encoder, where a FC layer is attached at the end of each FCDS block for performance improvement. }
    \label{fig: SCE network}
\end{figure*}

Most previous works focus on improving recovery performance and reducing model complexity for different compression ratios separately. This implies that multiple encoder-decoder pairs which are deployed at UEs and gNB are required for demands of distinct compression ratios. In \cite{CsiNet-SM}, a multi-rate CSI framework is proposed. In this framework, as illustrated in Fig. \ref{fig: CsiNet-SM}, the encoder in this framework contains $4$ outputs for $4$ different compression ratios. Note that the parameters of all layers in the encoder are shared except for the last fully-connected (FC) layer. This framework reduces the total number of parameters by considering the fact that features captured by convolutional layers for different compression ratios are similar. In this paper, we adopt a similar architecture but using a new design of encoder with fully convolutional layers and the proposed DC framework. We propose a new neural network called successive convolutional encoding network (SCEnet) where the model complexity can be largely reduced while maintaining the recovery performance.
To strike a balance between the performance and model complexity, we focus on reducing the complexity of encoder due to the limited computational budget at UEs. As for the encoder design, a fully-convolutional down-sizing block (FCDS) is introduced to reduce the input size by a factor of two. A FCDS block consists of $1\times7$, $1\times5$ and $1\times3$  convolutional layers with $2$ channels, respectively. Note that the stride lengths are all $1$ except for the last horizontal stride length in the last convolutional layer which is set as $2$ to reduce the input size by a factor of two. Fig. \ref{fig: SCE network} shows an example of a CSI feedback framework using $S$ (= 4 throughout this paper) FCDS blocks for dealing with $4$ compression ratios. Specifically, the output of $i$-th block with size of $K\cdot N_f/2^i$ represents the codeword with compression ratio = $2^i, i = 1,...,S$.
The decoder is designed individually for different compression ratios. For the $i$-th decoder, the codeword is first fed to a $K \cdot N_f$ FC layer, a $1 \times 3$ convolutional layer and activation function after reshaping for initial estimation. A RefineBlock \cite{CsiNet-SM} is followed for refinement. RefineBlock is a residual structure and consists of three $1 \times 3$ convolutional layers with $16$, $8$ and $1$ channels and activation functions. Then, it is followed by a $K \cdot N_f$ FC layer for generating the real/imaginary CSI estimate. To further improve the performance, we provide another version of SCEnet, called SCEnet+. We add an additional FC layer at the end of each FCDS block which provides extra non-linearity while maintaining the same output size.

The parameters of the SCEnet are optimized according to the following criterion:

\begin{equation}
    \Omega_{en}, \Omega_{de} = \arg\min \sum_i^D\sum_s^{S=4} 
    W_s \cdot \norm{\mathbf{H}_i - \widehat{\mathbf{H}}_{i,s}} ,
    \label{criterion}
\end{equation}
\begin{equation}
    \widehat{\mathbf{H}}_{i,1}, 
    \widehat{\mathbf{H}}_{i,2},
    \widehat{\mathbf{H}}_{i,3},
    \widehat{\mathbf{H}}_{i,4} = f_{de}(f_{en}(\mathbf{H}_i)),
\end{equation}
where $\Omega_{en}$, $\Omega_{de}$ denote the trainable parameters of encoder $f_{en}$ and decoder $f_{de}$. $D$ is the training data size. The hyper-parameters $\{W_1,W_2,W_3,W_4\}$ refers the setting $\{30/39, 6/39, 2/39, 1/39\}$ in \cite{CsiNet-SM}.

\section{Experimental Evaluations}
\subsection{Experiment Setup}
In our numerical test, we consider both indoor and outdoor cases. 
Using channel model software, 
we position a gNB of height equal to 20 m
at the center of a circular cell
with a radius of 30 m for indoor and 200 m for outdoor environment. 
We equip the gNB with a $8 \times 4 (N_H \times N_V)$ UPA for communication with 
single antenna UEs. UPA elements have half-wavelength uniform spacing.

\begin{figure}
    \centering
    \resizebox{3in}{!}{
    \includegraphics{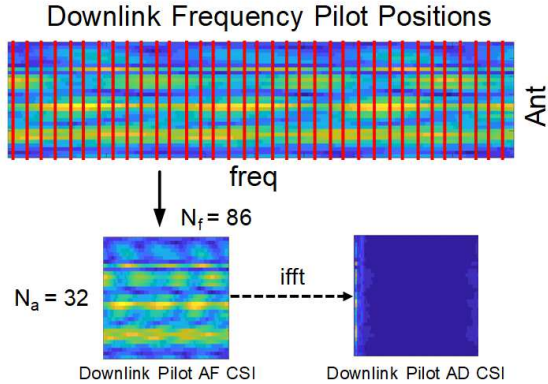}}
    \caption{Pilot placement illustration (Note that the red lines indicate the time-frequency resources to be placed pilot symbols. AF and AD stand for antenna-frequency and antenna-delay domains, respectively).}
    \label{fig: DLCSIPILOT}
\end{figure}

For our proposed model and other competing models, we set the number of epochs to $1,000$. We use batch size of $200$. For our model, we start with learning rate of $0.001$ before switching to
$5*10^{-4}$ after the $300$-th epoch. 
Using the channel simulator, We generate several indoor and outdoor datasets, each containing 100,000 random channels. 
57,143 and 28,571 random channels are for training and validation. 
The remaining 14,286 channels are 
test data for performance evaluation. 
For both indoor and outdoor, we use 
the QuaDRiGa simulator \cite{QuaDriGa} 
using the scenario features given in \textit{3GPP TR 38.901 Indoor} and \textit{3GPP TR 38.901 UMa} at 5.1-GHz and 5.3-GHz, and 300 and 330 MHz of UL and DL with LOS paths, respectively. Here, we assume UEs are capable of perfect channel estimation. For each data channel, we consider $N_f = 1024$ subcarriers with $15K$-Hz spacing and place $M_f = 86$ pilots with downsampling ratio $\text{DR}_f = 12$ as illustrated in the Fig. \ref{fig: DLCSIPILOT}. We set antenna type to \textit{omni}. We use normalized MSE as the performance metric
\begin{equation}
\frac{1}{D}\sum^{D}_{d=1}\norm{\widehat{\mathbf{H}}_{d} - \mathbf{H}_{d}}^2_\text{F} /\norm{\mathbf{H}_{d}}^2_\text{F},\label{NMSE1}
\end{equation}
where the number $D$ and subscript $d$ denote the total number and index of channel realizations, respectively. $\widehat{\mathbf{H}}_{d} \in \mathbb{C}^{{N_a}\times{N_f}}$ denotes the estimated DL CSI after padding zeros back to its full size. $\mathbf{H}_{d} \in \mathbb{C}^{{N_a}\times{N_f}}$ denotes the true DL CSI.
For comparison, other than the proposed models, SCEnet and SCEnet-Dense, we also include two multi-rate CSI feedback alternatives as follows:
\begin{itemize}
    \item \textbf{CsiNet-SM}: Fig. \ref{fig: CsiNet-SM} (a) shows the general architecture. Note that the same decoders as the proposed models for different compression ratios are adopted but the convolutional filter size is a two-dimensional version (i.e., (3,3), (5,5) and (7,7)). 
    \item \textbf{CsiNet-PM}: Fig. \ref{fig: CsiNet-SM} (b) shows the general architecture. Note that CsiNet-PM is a more compact model than CsiNet-SM but suffers slightly performance degradation in general.
\end{itemize}
Note that all the above multi-rate CSI frameworks focus on the compression ratios $= 2$, $4$, $8$ and $16$. 
\subsection{SCEnet vs. SCEnet-Dense}
\begin{figure}
    \centering
    \resizebox{3.4in}{!}{
    \includegraphics{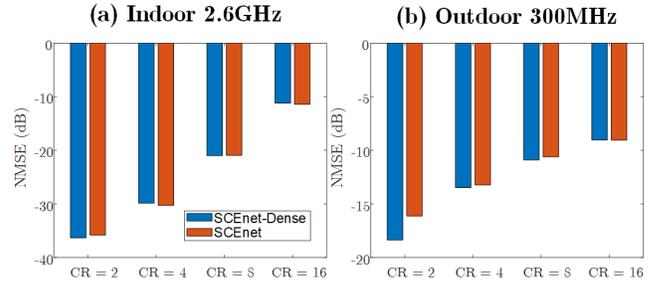}}
    \caption{NMSE performance at different compression ratios for SCEnet and SCEnet-Dense in indoor and outdoor scenarios.}
    \label{fig: DiffStructure}
\end{figure}

Figs. \ref{fig: DiffStructure} (a) and (b) demonstrate the NMSE performance of the two proposed models at different compression ratios for indoor and outdoor scenarios, respectively. We first discover that a decent performance improvement by adding extra FC layers if the number of RefineBlocks is small. However, we can also find that the additional FC layers do not contribute obvious performance improvement if the number of RefineBlock is $5$. Previous works usually adopt FC layer at encoder to provide enough non-linearity. Yet, this result implies that it is not necessary to adopt the FC at encoder if a powerful decoder is utilized. As compared to UEs, gNB usually does not suffer from computational and storage limitations. Therefore, it is more reasonable to use a simple encoder with the help of a complicated decoder.

\subsection{Comparison with SOTAs and scalibility of SCEnet}
\begin{figure}
    \centering
    \resizebox{3.4in}{!}{
    \includegraphics{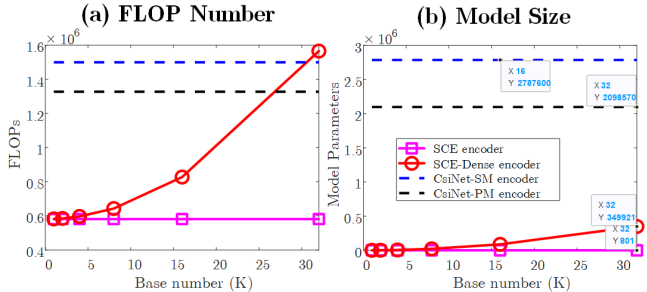}}
    \caption{Comparison of (a) numbers of FLOPs and (b) model parameters at various base numbers, $K$. Note that $K = 32$ for CsiNet-SM and CsiNet-PM. ($K = 1,2,4,8,16,32$)}
    \label{fig: FLOP_PARA}
\end{figure}

Table \ref{tab:my-table} shows the NMSE performance at different compression ratios and base number ($K$) for SCEnet, CsiNet-SM and CsiNet-PM in both indoor and scenarios. Fig. \ref{fig: FLOP_PARA} shows the FLOP number and model size of encoder at base number ($K$) for SCEnet, SCEnet-Dense, CsiNet-SM and CsiNet-PM. We can first find that the SCEnet when $K = 32$ generally outperforms CsiNet-SM and CsiNet-PM while requiring far less FLOP number and storage at UE side. On the other hand, by introducing the DC framework, when we utilize a base number $K$ which can divide $N_a$, we can enjoy lower complexity and less required storage at UEs while suffering slight performance degradation. Moreover, by adopting $K = 2$, SCEnet can be a universal CSI feedback framework which can be applied to different number of antenna ports (according to the 3GPP specification, 2, 4, 8, 16, 32 are possible antenna port number).

\begin{figure}
    \centering
    \resizebox{3.4in}{!}{
    \includegraphics{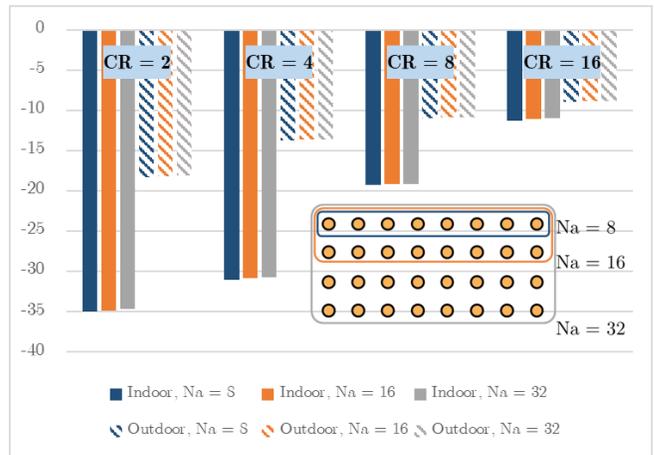}}
    \caption{NMSE performance of SCEnet for arrays with different numbers of antennas. Note that the when $N_a = 8, 16, 32$ we use the first, first two and all rows of the UPA as different arrays.}
    \label{fig: DiffAnt}
\end{figure}

Fig. \ref{fig: DiffAnt} shows the NMSE performance at different compression ratios, propagation channels, and number of antennas being used for SCEnet. We can find that there are no obvious performance difference when considering arrays with different numbers of antennas. This result demonstrates the scalibility of the proposed DC framework.

\begin{table}[]
\centering
\caption{NMSE performance of the CsiNet-SM and SCEnet for different base numbers ($K$).}
\label{tab:my-table}
\begin{tabular}{|c|c|cccccc|}
\hline
\multirow{2}{*}{CR} & \multirow{2}{*}{Scen} & \multicolumn{4}{c|}{SCEnet}                                                                                                                                    & \multicolumn{1}{c|}{\begin{tabular}[c]{@{}c@{}}CsiNet\\ -SM\end{tabular}} & \begin{tabular}[c]{@{}c@{}}CsiNet\\ -PM\end{tabular} \\ \cline{3-8} 
                    &                       & \multicolumn{1}{c|}{K=1}   & \multicolumn{1}{c|}{K=2}                  & \multicolumn{1}{c|}{K=8}                  & \multicolumn{3}{c|}{K=32}                                                                                                                                                    \\ \hline
\multirow{2}{*}{2}  & Ind.                  & \multicolumn{1}{c|}{-33.3} & \multicolumn{1}{c|}{{-34.8}} & \multicolumn{1}{c|}{\textbf{-35.4}}                & \multicolumn{1}{c|}{-33.0}                & \multicolumn{1}{c|}{-31.6}                                                & -31.7                                                \\ \cline{2-8} 
                    & Out.                  & \multicolumn{1}{c|}{-17.4} & \multicolumn{1}{c|}{-18.2}                & \multicolumn{1}{c|}{{ \textbf{-21.2}}} & \multicolumn{1}{c|}{{\textbf{-21.2}}} & \multicolumn{1}{c|}{-20.0}                                                & -19.8                                                \\ \hline
\multirow{2}{*}{4}  & Ind.                  & \multicolumn{1}{c|}{-28.6} & \multicolumn{1}{c|}{{\textbf{-31.0}}} & \multicolumn{1}{c|}{-30.0}                & \multicolumn{1}{c|}{-27.2}                & \multicolumn{1}{c|}{-26.8}                                                & -26.4                                                \\ \cline{2-8} 
                    & Out.                  & \multicolumn{1}{c|}{-13.8} & \multicolumn{1}{c|}{-14.0}                & \multicolumn{1}{c|}{-16.6}                & \multicolumn{1}{c|}{{ \textbf{-18.1}}} & \multicolumn{1}{c|}{-16.2}                                                & -15.2                                                \\ \hline
\multirow{2}{*}{8}  & Ind.                  & \multicolumn{1}{c|}{-21.2} & \multicolumn{1}{c|}{-21.3}                & \multicolumn{1}{c|}{-21.2}                & \multicolumn{1}{c|}{{ \textbf{-21.6}}} & \multicolumn{1}{c|}{-21.2}                                                & -20.2                                                \\ \cline{2-8} 
                    & Out.                  & \multicolumn{1}{c|}{-10.7} & \multicolumn{1}{c|}{-10.9}                & \multicolumn{1}{c|}{-12.6}                & \multicolumn{1}{c|}{{ \textbf{-14.0}}} & \multicolumn{1}{c|}{-12.8}                                                & -11.7                                                \\ \hline
\multirow{2}{*}{16} & Ind.                  & \multicolumn{1}{c|}{-12.2} & \multicolumn{1}{c|}{-12.8}                & \multicolumn{1}{c|}{-12.4}                & \multicolumn{1}{c|}{{ \textbf{-13.5}}} & \multicolumn{1}{c|}{-13.1}                                                & -12.6                                                \\ \cline{2-8} 
                    & Out.                  & \multicolumn{1}{c|}{-8.9}  & \multicolumn{1}{c|}{-8.9}                 & \multicolumn{1}{c|}{-9.7}                 & \multicolumn{1}{c|}{{ \textbf{-11.0}}} & \multicolumn{1}{c|}{-10.6}                                                & -9.6                                                 \\ \hline
\end{tabular}
\end{table}

\section{Conclusions}
This work proposes a divide-and-conquer (DC) CSI feedback framework which is universal for different number of antenna ports defined in 3GPP specification, but also lowers the computational complexity and model storage requirement at UE side by allowing UE paralelly feedbacks segmented DL CSI with respect to the low correlation between antennas (or antenna ports). The framework consists of a multi-rate successive convolutional encoder (SCE) without any FC layer which usually is the major reason causing a fat model. In numerical results, the proposed framework generally outperform than the SOTAs, CsiNet-SM and CsiNet-PM, while requiring much lower computation and storage requirements for UEs, which usually have tight resource budgets.
\section{Acknowledgement}
The authors would like to acknowledge Mason del Rosario for his useful discussions which helped the authors better understand of pilot placement and channel truncation.
\bibliography{references.bib}
\bibliographystyle{IEEEtran}
\end{document}